\documentclass[%
 reprint,
 amsmath,amssymb,
 aps,
showkeys
]{revtex4-2}

\usepackage{graphicx}
\usepackage{dcolumn}
\usepackage{bm}
\usepackage{hyperref}

\begin{document}

\preprint{}

\title{Maximal Transcendentality of the Double-Scaled PCM}

\author{Evgeny Sobko}%
\email{corresponding author:  evgenysobko@gmail.com}
\affiliation{London Institute for Mathematical Sciences, Royal Institution, London, W1S 4BS, UK}%


\begin{abstract}
We prove, to all orders, maximal transcendentality of the strongly coupled large-\(N\) Principal Chiral Model in the double-scaling regime introduced in our earlier work. We also prove that, after a natural shift of the coupling constant, the coefficients of the vacuum-energy expansion are expressed purely as polynomials in odd zeta values with rational coefficients. The first 35 explicitly computed orders reveal further number-theoretic regularities, pointing to hidden structure beyond maximal transcendentality.

\end{abstract}

\maketitle

The principle of maximal transcendentality originated in the work of Kotikov and Lipatov on DGLAP/BFKL dynamics in \(\mathcal{N}=4\) super-Yang--Mills theory (SYM)~\cite{Kotikov:2002ab}, where twist-two anomalous dimensions were observed to follow from the corresponding QCD results by retaining only the highest-transcendentality terms. The underlying notion is that of transcendental weight\cite{Duhr:2014woa}, a grading that reflects the complexity of the special functions and numbers appearing in quantum field theory; a related notion, uniform transcendentality, refers to the case in which all terms at a fixed perturbative order have the same weight. These ideas have since become powerful organizing principles for perturbative expansions in \(\mathcal{N}=4\) SYM, from anomalous dimensions \cite{Marboe:2014gma} and Wilson loops to scattering amplitudes and form factors, and underlie both modern amplitude bootstraps and the canonical differential-equation approach to Feynman integrals~\cite{Henn:2013pwa}. Similar near-uniform or highly constrained transcendental structures also appear in conformal fishnet theories\cite{Gurdogan:2015csr}, obtained as double-scaled limits of \(\gamma\)-deformed \(\mathcal{N}=4\) SYM, and in tree-level closed-string amplitudes, whose low-energy expansion is governed by single-valued multiple zeta values~\cite{Stieberger:2013wea,Brown:2019wna}.

Despite their importance, maximal and uniform transcendentality remain, in most settings, empirical observations rather than theorems. They are usually observed within perturbation theory, through diagrammatic calculations or bootstrap data. This diagrammatic origin has led to deep connections with number theory, periods, motives, and the cosmic Galois group~\cite{Brown:2015fyf}. At the same time, rigorous all-orders proofs for specific physical observables remain rare, and strong-coupling analogues are almost absent. Moreover, most known examples arise in supersymmetric theories, primarily \(\mathcal{N}=4\) SYM and ABJM.

The strongly coupled \(SU(N)\) Principal Chiral Model (PCM) ~\cite{Polyakov:1975rr,Polyakov:1977vm} provides a sharply different setting. It has none of the usual ingredients associated with maximal or uniform transcendentality, except integrability\cite{Polyakov:1984et,Wiegmann:1984ec}: no supersymmetry, no conformal symmetry, and no diagrammatic perturbative origin. Instead, the PCM is often viewed as one of the closest two-dimensional integrable analogues of four-dimensional QCD: it is asymptotically free, dynamically generates a mass gap, and admits a ’t Hooft large-\(N\) expansion. The latter suggests a possible dual string description at strong coupling, where planar diagrams become dense. In earlier work~\cite{Kazakov:2019laa}, the author, together with V.~Kazakov and K.~Zarembo, introduced a double-scaling (DS) limit combining large \(N\) and strong coupling. The resulting theory exhibited string-like behaviour reminiscent of the \(c=1\) matrix model~\cite{Kazakov:1988ch}, whose DS limit~\cite{Brezin:1989ss,Gross:1990ay,Ginsparg:1990as} is dual to a 2D string theory, motivating a conjectural 3D string dual.

In this Letter, we prove maximal transcendentality of the DS PCM to all orders in the inverse coupling: every coefficient of the vacuum energy expansion has a maximally transcendental weight equal to its order. Moreover, by uncovering a hidden gauge symmetry and performing a simple shift of the coupling constant, we prove that the expansion is expressed entirely in terms of polynomials in odd zeta values with rational coefficients.

The mechanism behind this transcendental structure is distinct from known examples: it comes from a subtle interplay between the special structure of the DS limit and the recursive nature of the Wiener-Hopf (WH) solution, rather than from the usual diagrammatic techniques. Related WH-based methods have recently been used to generate weak-coupling asymptotic expansions in various integrable models \cite{Volin:2009wr, Marino:2019eym,DiPietro:2021yxb,Abbott:2020qnl}. In those cases, however, the series are generic, with coefficients of mixed transcendentality and can be traced back to their diagrammatic origin. Thus, although the technical machinery is similar, its role here is qualitatively different: combined with the double-scaled structure, it does not merely generate an asymptotic series, but enforces a hidden maximally transcendental organization with non-diagrammatic realization at strong coupling.

The first 35 explicitly computed orders exhibit further number-theoretic regularities. Each coefficient appears to contain all monomials in odd zeta values of the corresponding weight, always with positive rational coefficients, hinting at a possible combinatorial or volume interpretation. In addition, the action of derivatives with respect to odd zeta values on the vacuum-energy coefficients points to a hidden, possibly motivic, structure beyond maximal transcendentality.


\section{Double-scaled PCM}

In this section we briefly review the vacuum Bethe-ansatz equation for the double-scaled PCM derived in \cite{Kazakov:2019laa}.

The \(SU(N)\) Principal Chiral Model is defined by the action
\begin{gather}
S=\frac{N}{\lambda_0}\int d^2x\, \mathop{\mathrm{tr}} D_\mu g^\dag D^\mu g,
\end{gather}
where the field \(g(x)\in SU(N)\). Its spectrum consists of \(N-1\) massive particles transforming in bi-fundamental representations of the global \(SU(N)\times SU(N)\) symmetry. The lightest particle acquires its mass through dimensional transmutation, while the remaining particles may be viewed as bound states with masses
\begin{gather}
m_a=m\frac{\sin\frac{\pi a}{N}}{\sin\frac{\pi}{N}},
\quad a=1,\ldots,N-1 .
\end{gather}

A finite density of particles is introduced by gauging the global \(SU(N)\times SU(N)\) symmetry with constant chemical potentials \(D_0 g=\partial_0 g-\frac{i}{2}(Hg+gH)\), \(D_1 g=\partial_1 g \).

The PCM is an integrable quantum field theory, and for the special choice of \(H=\mathop{\mathrm{diag}}(h_1,h_2-h_1,\ldots,h_{N-1}-h_{N-2},-h_{N-1})\) \cite{Fateev:1994ai,Kazakov:2019laa}:
\begin{gather}
h_a=h\frac{\sin\frac{\pi a}{N}}{\sin\frac{\pi}{N}},
\quad a=1,\ldots,N-1 .
\end{gather}
the finite-density Bethe-ansatz equations for the vacuum energy simplify drastically \cite{Fateev:1994ai,Kazakov:2019laa}. In the thermodynamic limit they reduce to a single integral equation for a pseudoenergy, depending on the rapidity \(\theta\), with finite support \(\theta\in(-B,B)\). The DS limit \cite{Kazakov:2019laa} is defined as 
\begin{gather}
B\rightarrow0,\qquad N\rightarrow\infty,\qquad b=BN\ \ - \ \text{fixed}.
\end{gather}

his DS regime the Bethe-ansatz equation takes the form
\begin{gather}
\epsilon(t)+\int\limits_{-b}^b \mathcal{K}(t-t')\epsilon(t')\,dt'
=\delta(b)-\frac{t^2}{2},
\label{DSinCoord}
\end{gather}
where \(\mathcal{K}(t)=\frac{1}{\pi^2}
(2\psi(1)-\psi(1-\frac{i t}{\pi})
-\psi(1+\frac{i t}{\pi}))\), \(\epsilon(t)\) is the pseudoenergy and the boundary condition \(\epsilon(\pm b)=0\) fixes the constant \(\delta(b)\).

The vacuum energy is given as:
\begin{gather}\label{DSenergy}
E=\frac{1}{4\pi^3}\int\limits_{-b}^b dt\, \epsilon(t)\,.
\end{gather}
This determines \(E\) parametrically as a function of \(b\). 

For small \(b\), the integral operator in Eq.~\eqref{DSinCoord} is contractive, allowing a simple iterative solution; see appendix \ref{AppendixSmallb} for details. This gives
\begin{gather}
E\underset{b\to 0}{=}\frac{1}{6}\frac{b^3}{\pi^3}
-\frac{4\zeta(3)}{9\pi}\frac{b^6}{\pi^6}
+\frac{16\zeta(5)}{15\pi}\frac{b^8}{\pi^8}
+\frac{32\zeta(3)^2}{27\pi^2}\frac{b^9}{\pi^9}
+\cdots
\end{gather}

In this paper we focus on the large-\(b\) regime, where the pseudoenergy
admits the bulk expansion:
\begin{gather}\label{DSBulkExp}
\epsilon(t)=\sum\limits_{l=0}^\infty  \sum\limits_{j=0}^\infty \beta_{\frac{1}{2}-l,l-j}\, b^{l-j}(b^2-t^2)^{\frac{1}{2}-l}
\end{gather}
with \(\beta_{\frac{1}{2},0}=1\). Expanding near the edge \(t=-b\) and Fourier transforming on the half-line gives:
\begin{gather}\label{EpsExp}
\epsilon(k)=\sum\limits_{l=0}^\infty  \sum\limits_{j=0}^\infty \beta_{\frac{1}{2}-l,l-j}\, b^{l-j} \rho_l(k)
\end{gather}
\begin{gather}
\rho_l(k)=\sum\limits_{n=0}^\infty \Gamma(\frac{3}{2}+n-l)\frac{(l-1/2)_n}{n! \ 2^{n+l-1/2}}\frac{(-i k)^{-3/2+l-n}}{b^{l+n-1/2}}
\end{gather}
The Wiener-Hopf matching of the bulk and boundary asymptotics yields a closed linear system for the coefficients \(\beta_{\frac{1}{2}-l,l-j}\) :
\begin{gather} \label{WHeqn}
\epsilon(k)=G_+(k) \mathop{\mathrm{res}}_{p=0} \left[\frac{1}{k-p}G^{-1}_{+}(p) \epsilon(p)\right]
\end{gather}
where \(G_{\pm}(p)\) is the Wiener-Hopf factorisation 
\begin{gather}
G_+(p)=\frac{1}{\sqrt{\pi}p^\frac{1}{2}}\frac{\Gamma(\frac{1}{2}-i\frac{p}{2})}{\Gamma(1-i\frac{p}{2})}
\end{gather}
of the kernel $\mathcal{K}(p)=\coth\frac{\pi |p|}{2}=(G_+(p)G_-(p))^{-1}$, where \(|p|\) is implied  as \(\sqrt{p+i0}\sqrt{p-i0}\).

In the DS limit, the energy receives contributions only from the two modes \((b^2-t^2)^{\pm\frac{1}{2}}\) \cite{Kazakov:2019laa}. Thus, after solving \eqref{WHeqn}, one obtains:
\begin{gather}\label{energy}
E=\frac{b^2}{8\pi^2}(1+2\sum\limits_{j=0}^\infty \beta_{-\frac{1}{2},1-j}\, b^{-j-1})
\end{gather}

\section{Maximal transcendentality and the odd-zeta structure}\label{sec:MT}
The Wiener-Hopf factor 
 \(G_+(p)=p^{-\frac{1}{2}}R_+(p)\) can be written as:
\begin{gather}
R_+(p)=e^{i p \log 2}U(p)V(p)= \sum\limits_{n=0}^\infty r_n (ip)^n,\label{R_+}\\
U(p)=\exp\,
\sum_{l=1}^{\infty}
\frac{(i p)^{2l}}{2l}\eta(2l)
,\label{U}\\
V(p)=\exp\,
\sum_{l=1}^{\infty}
\frac{(i p)^{2l+1}}{2l+1}\eta(2l+1).
\label{V}
\end{gather}
Here \(\eta(n)=(1-2^{1-n})\zeta(n)\) is the Dirichlet eta function, written in terms of the Riemann zeta function at integer arguments. The coefficients \(r_n\) are naturally expressed through the complete exponential Bell polynomials: 
\begin{gather}
r_n=
\frac{1}{n!}\,
B_n\!\left(
\log 2,\,
1!\eta(2),\,
\dots,\,
(n-1)!\eta(n)
\right).\label{coefsn}
\end{gather}

We also introduce the analogous expansion for the inverse factor \(R_+^{-1}(p)\):
\begin{gather}
R_+^{-1}(p)=\sum\limits_{n=0}^\infty \bar{r}_n(ip)^n
\end{gather}

We now introduce the notion of transcendental weight in the minimal form needed for our consideration; for a broader discussion, see \cite{Duhr:2014woa}. In what follows we consider only polynomials in \(\log 2\) and zeta values \(\zeta(n)\), with rational coefficients. Rational numbers have weight zero, \(w(p/q)=0\), while \(w(\log 2)=1\) and \(w(\zeta(n))=n\). The weight is additive on products, \(w(M_1M_2)=w(M_1)+w(M_2)\); for example, \(w(\log2\, \zeta(3)^9)=28\). Even zeta values can equivalently be written as rational multiples of powers of \(\pi\), with \(w(\pi)=1\): \(\zeta(2k)=(-1)^{k+1}\,
\frac{B_{2k}}{2(2k)!}\,
(2\pi)^{2k}\), where \(B_{2k}\) are Bernoulli numbers.

A sum is uniformly transcendental if all its terms have the same weight. The form of Bell polynomials \eqref{coefsn} then imply that \(r_n\) and \(\bar r_n\) are uniformly transcendental of weight \(n\):
\begin{gather}
w(r_n)=w(\bar r_n)=n.
\end{gather}
Equation \eqref{WHeqn} can be solved iteratively starting with the initial conditions:
\begin{gather}
\beta_{\frac{1}{2},0}=1; \ \ \ 
\beta_{\frac{1}{2},-l}=0, \ \  l\geq1. 
\end{gather}
We organize beta coefficients by index \(ind(\beta_{1/2-l,l-j})=l+j\). The WH recursion is triangular with respect to this index. More precisely, at step \(N+1\) we already have all coefficients with index \(ind\leq N\) and coefficients with index \(ind=N+1\) are determined by the following linear system of \(N+1\) equations \(s\in {0,1,\dots,N}\) :
\begin{gather}\label{WH_iteration}
\sum_{j=0}^{s} A_{sj}^{(N)}\,\beta_{j-N-\frac{1}{2},N+1-2j}=B_s^{(N)},
\end{gather}
where
\begin{gather}
A_{sj}^{(N)}
=
\frac{(-1)^{N-s}\left(N+\frac12-j\right)_{s-j}}
{(s-j)!\,2^{N+s+1-2j}},\quad j\leq s\notag\\
\qquad A_{sj}^{(N)}=0, \quad j>s
\end{gather}
and
\begin{gather}
B_s^{(N)}=\sum _{h=0}^s \sum _{p=h}^{s} (-1)^{p+1}\, \bar{r}_{p-h}\, r_{h+N-s+1}\notag\\
\times\frac{\Gamma(p+\frac{3}{2})}{\Gamma(s+\frac{1}{2}-N)}\sum _{l=0}^{s-p}\frac{(l-\frac{1}{2})_{p+l}}{(p+l)!\, 2^{p+2l}}\, 
\beta_{\frac{1}{2}-l,p+2 l-s}
\end{gather}

We can now readily prove the following
\par\medskip
\textbf{Proposition 1}
All coefficients \(\beta_{\frac{1}{2}-l,l-j}\) 
belong to the \(w\)-graded ring \(\mathbb Q\big[\log 2,\pi^2, \zeta(3),\zeta(5),\ldots\big]\), and are uniformly transcendental of weight \(w(\beta_{\frac{1}{2}-l,l-j})=l+j\). 
\par\medskip
\textbf{Proof} is by induction on the  index \(ind=l+j\). The initial condition \(\beta_{\frac12,0}=1\) has weight zero. Assume the claim for all coefficients with \(ind\leq N\).  Coefficients with \(ind=N+1\) are defined by the system \eqref{WH_iteration}. Every element \(B_s^{(N)}\) on the right-hand side  is a rational linear combination of products of \(r\), \(\bar{r}\) and \(\beta\) coefficients with \(ind\leq N\). The weights: \(w(r_{h+N-s+1})=h+N-s+1\), \(w(\bar{r}_{p-h})=p-h\) and \(w(\beta_{\frac{1}{2}-l,p+2 l-s})=s-p\), so every term has total weight \(N+1\):
\begin{gather}
w(B_s^{(N)})=(p-h)+(h+N-s+1)+(s-p)=N+1.\notag
\end{gather}
The matrix \(A^{(N)}\) and its inverse \eqref{InverseA} are also rational, so we get that all beta coefficients with index \(ind=N+1\) have weight \(w=N+1\). Because \(\beta_{\frac{1}{2},0}=1\) and \(r_n,\bar{r}_n\in\mathbb Q\big[\log 2,  \zeta(2),\zeta(3),\ldots\big]\) we get \(\beta_{\frac{1}{2}-l,l-j}\in \mathbb Q\big[\log 2,\pi^2, \zeta(3),\zeta(5),\ldots\big]\) and the claim follows \(\blacksquare\)
\par\medskip
In particular, this gives \(w(\beta_{-\frac{1}{2},1-j})=j+1\) for coefficients in the energy's expansion \eqref{energy}.

\par\medskip

We now analyze how the \(\beta\) coefficients depend on \(\pi^2\). The only source of \(\pi^{2k}\) is the even part \(U(k)\) \eqref{U}  of \(R_+(k)\). The direct calculation gives that 
\begin{gather}\label{p2Mult}
k^2 \rho_l(k)=2l(2l-1)\rho_{l+1}(k)-(4l^2-1)b^2\rho_{l+2}(k)
\end{gather}
with the special case \(l=0\) where \(\rho_1\) is absent
\begin{gather}\label{gap}
k^2 \rho_0(k)=b^2\rho_{2}(k)
\end{gather}
This means that multiplication by any even power \(k^{2m}\) or more generally by any series in even powers of \(k\) is closed on the space of functions \(\{\rho_l(k)\}\). We therefore define \(\tilde\epsilon(k)\) by dividing out the even part \(U(k)\):
\begin{gather}
\tilde{\epsilon}(k)=U^{-1}(k)\epsilon(k)
\end{gather}
which preserves the expansion \eqref{EpsExp} and \(w\)-grading, though now with new coefficients \(\tilde{\beta}\). Indeed, the inverse even part \(U(k)^{-1}\) has an expansion :
\begin{gather}
U(k)^{-1}=1+\bar{u}_2k^2+\bar{u}_4k^4+...
\end{gather}
with \(w(\bar{u}_{2k})=2k\). The multiplication by \(\bar{u}_2k^2\) acts as 
\begin{gather}
\bar{u}_2k^2\epsilon(k)=\sum\limits_{l=2}^\infty  \sum\limits_{j=1}^\infty 2(l-1)(2l-3)\bar{u}_2\beta_{\frac{3}{2}-l,l-j}\, b^{l-j} \rho_l(k)\notag\\
-\sum\limits_{l=2}^\infty  \sum\limits_{j=0}^\infty (4(l-2)^2-1)\bar{u}_2\beta_{\frac{5}{2}-l,l-2-j}\, b^{l-j} \rho_l(k)\label{MultByk^2}
\end{gather}
so we see that it has the same form and coefficients have the same weight as in \(\epsilon(k)\):
\begin{gather}
w(\bar{u}_2\beta_{\frac{3}{2}-l,l-j})=w(\bar{u}_2\beta_{\frac{5}{2}-l,l-2-j})=j+l
\end{gather}
A similar expansion holds for the higher powers \(\bar{u}_{2m}k^{2m}\). 

Crucially, \eqref{p2Mult} and  \eqref{gap} imply that multiplication by \(U^{-1}(k)\) leaves \(\rho_0\) and \(\rho_1\) untouched, so that:
\begin{gather}
\tilde{\beta}_{\frac{1}{2},-j}=\beta_{\frac{1}{2},-j},\ \ \tilde{\beta}_{-\frac{1}{2},1-j}=\beta_{-\frac{1}{2},1-j}
\end{gather}
The original equation \eqref{WHeqn} can be rewritten as 
\begin{gather}
\tilde{\epsilon}(k)=e^{i k \log 2}\, V(k) \mathop{\mathrm{res}}_{p=0} \left[\frac{e^{-i p \log 2}}{k-p}V^{-1}(p) \tilde{\epsilon}(p)\right]
\end{gather}
and solved iteratively in the same way as described above for the full \(R_{+}\). Hence the coefficients \(\tilde{\beta}\) are polynomials in odd zetas and \(\log 2\) only. In particular, the energy \eqref{energy}, expressed solely through the \(\beta_{-\frac{1}{2},1-j}\), is also independent of even zetas.

Going back from \(\tilde{\epsilon}(k)\) to \(\epsilon(k)\) and using the analogue of \eqref{MultByk^2} for general \(u_{2m}k^{2m}\) it is straightforward to make a more general statement about maximal power of \(\pi\) in all original \(\beta\) coefficients (\(l\geq1\)):
\par\medskip
\textbf{Proposition 2}
\begin{gather}
\deg_{\pi}(\beta_{1/2-l,l-j})\leq\min\left(2(l-1),2\left[\frac{j+l}{2}\right]\right)\label{degpi}
\end{gather}
Finally, we note that our WH problem is formulated on the half-line \((-b,\infty)\) and the energy \(E\) can be reexpanded around any shifted point \(-b\rightarrow-b+c\). \(E(\epsilon(k,b))=E(e^{i k c}\, \epsilon(k,b-c))\). Choosing \(c=\log 2\) cancels \(e^{i k \log2}\) in \(R_+\). This, in particular, implies that 
\begin{gather}
\partial_{\, \log2}\, \beta_{-\frac{1}{2},-j}
= (j-1)\,\beta_{-\frac{1}{2},1-j}\label{log2der}
\end{gather}
where we treat \(\log2\) as a variable and \(\beta\)'s are polynomials in it.

Combining this with Propositions 1 and 2  we arrive at the main result of our paper:
\par\medskip
\textbf{Theorem}

The energy \eqref{energy} of the DS PCM in the limit  \(b=\bar{b}+\log2\rightarrow \infty\)  is given by the expansion \(E=\frac{\bar{b}^2}{8\pi^2}(1+2\sum\limits_{j=0}^\infty \beta^V_{-\frac{1}{2},1-j}\, \bar{b}^{-j-1})\) with uniformly transcendental coefficients \(\beta^V_{-\frac{1}{2},1-j}\in \mathbb{Q}\big[ \zeta(3),\zeta(5),\ldots\big]\) and the weight \(w(\beta^V_{-\frac{1}{2},1-j})=j+1\). Explicitly, coefficients \(\beta^V_{-\frac{1}{2},1-j}\) are given by the same iterative solution of \eqref{WH_iteration} with \(R_+\) replaced by \(V\).
\par\medskip
We have verified this result, together with all the statements above — in particular \eqref{degpi} and \eqref{log2der} — explicitly, by computing the first 22 orders in \(1/b\).

The first nine orders in the \(1/\bar b\) expansion read:
\begin{gather}
\frac{8\pi^2}{\bar{b}^2}E(b)=1+\frac{3 \zeta (3)}{16 \bar{b}^3}+\frac{135 \zeta (5)}{1024 \bar{b}^5}
+\frac{81 \zeta (3)^2}{1024 \bar{b}^6}+\\
\frac{14175 \zeta (7)}{32768 \bar{b}^7}+\frac{27135 \zeta (5) \zeta (3)}{32768 \bar{b}^8}+\frac{189 \left(6912 \zeta (3)^3+74375 \zeta (9)\right)}{4194304 \bar{b}^9}\notag
\end{gather}
Further terms are given in the Supplemental Material.
\par\medskip

We conclude this section with a remark on the mechanism of \(\pi\) cancellation.  Whenever we are interested in physical observables expressible in terms of \(\{\beta_{-1/2,1-l}\}\), such as the  energy, we can consider a general gauge transformation \(\epsilon(k)\rightarrow g(k)\epsilon(k)\) of the form \(g(k)=\exp (\sum\limits_{n=1}^\infty (ik)^{2n} c_{2n} )\) and it leaves  \(\{\beta_{-1/2,1-l}\}\) invariant. All these transformations form a multiplicative abelian gauge group \(\mathcal G_{\mathrm{even}}\). Moreover, if all parameters carry weight \(w(c_{2n})=2n\), the gauge transformation preserves the grading of all coefficients \(\beta\).  This group \(\mathcal G_{\mathrm{even}}^t\) of transcendentality preserving transformations is a dense subgroup of \(\mathcal G_{\mathrm{even}}\), and \(U(k)\in\mathcal G_{\mathrm{even}}^t\). More generally, the functions \(g(k)=\exp (\sum\limits_{n=m}^\infty (ik)^{2n} c_{2n} )\) form the group \(\mathcal G^{(m)}_{\mathrm{even}}\subset \mathcal G_{\mathrm{even}}\) (and the grading-preserving analogue, \(\mathcal G^{(m),t}_{\mathrm{even}}\)), of transformations leaving \(\{\beta_{-1/2,\#},...,\beta_{1/2-m,\#}\}\) invariant. 

In other words, the \(\pi\) cancellation is not a feature of the particular functional form of \(U\), but a manifestation of a hidden transcendentality-preserving gauge symmetry of the DS PCM.

\section{Observations} 
\label{sec:Observations}

Using the result of the Theorem from the previous section, we computed (see the Supplemental Material) the vacuum expansion to 35th order in \(1/\bar{b}\) and observed several notable properties. In contrast to the rigorous results of the previous section, the following three statements are empirical observations drawn from this expansion, which, as we discuss, raise new intriguing questions.

\par\medskip
1) \textbf{Full span}. Each coefficient \(c_{j+1}=2\beta^V_{-\frac{1}{2},1-j}\) contains all possible monomials in odd zetas of the corresponding weight \(w=j+1\). The dimension \(d_w\) of the space of weight-\(w\) monomials is given by the Hilbert series:
\begin{gather}
H(t)\;=\;\sum_{w\geq 0} d_w\, t^w
\;=\;\prod_{n=1}^{\infty}\frac{1}{1-t^{2n+1}}.
\end{gather}
This property is expected, since the Taylor expansion of \(V(k)\) contains all possible monomials inside the Bell polynomials, and the recursive step of the iterative WH solution thus involves multiplication by all possible monomials; a complete proof would nonetheless require showing that no accidental cancellations occur.
\par\medskip

2) \textbf{Positivity}. The coefficients in front of all monomials in \(c_j\) are strictly positive. This might point to a combinatorial or volume interpretation of these coefficients. We note, however, that this does not hold for the general coefficients \(\beta^V_{\frac{1}{2}-l,l-j}\) - we have checked that these can contain monomials of differing signs.

\par\medskip

3) \(\mathbf{\bm{\partial}_{\bm{\zeta}_{2k+1}} c_n\approx \bm{\lambda} c_{n-2k-1}}\). For all 35 coefficients and for all derivatives with respect to odd zeta values, we find
\(\partial_{\zeta(2k+1)} c_n\approx \lambda\, c_{n-2k-1}\), where both sides are regarded as vectors in the basis of odd-zeta monomials. The collinearity is unexpectedly strong: the typical deviation is of order \(10^{-3}\). We quantify it by the sine of the angle \(\theta\) between the coefficient vectors of \(\partial_{\zeta(2k+1)} c_n\) and \(c_{n-2k-1}\). For example, in the basis \(\{\zeta(3)^3,\zeta(9)\}\), the coefficient vectors of \(\partial_{\zeta(3)}c_{12}\) and \(c_9\) are \(a=\left(\frac{1754703}{131072},\frac{19187634375}{134217728}\right)\), 
\(b=\left(\frac{5103}{16384},\frac{14056875}{4194304}\right)\),
respectively, giving \(\sin\theta_{a,b}=7.0\times10^{-4}\). Over all pairs we find \(\max \sin\theta=0.0011\), showing that the \(c_n\) are within \(0.1\%\) of being "eigenvectors" of the derivatives \(\partial_{\zeta(2k+1)}\), in this weight-lowering sense. This observation may point to a hidden motivic structure.

\section{Discussion}
\label{sec:Discussion}

In this Letter we proved maximal transcendentality of the double-scaled PCM and showed that, in a natural scheme, the vacuum-energy expansion is expressed entirely through polynomials in odd zeta values with rational coefficients. This provides an exceptional all-orders example of such a structure in a non-supersymmetric, non-conformal, strongly coupled QFT. The mechanism is also novel: it rests on the interplay between the special properties of the DS limit and the recursive Wiener--Hopf structure, and is qualitatively different from the standard diagrammatic origin of transcendentality.

The cancellation of powers of \(\pi\) relies on the hidden gauge symmetry \(\mathcal G_{\mathrm{even}}\) ( \(\mathcal G_{\mathrm{even}}^t\)). It would be very interesting to understand the physical role of this symmetry, particularly in the context of conjectural dual string description~\cite{Kazakov:2019laa,Kazakov:2023imu}, where it may encode part of the quantum-gravitational degrees of freedom.

Our results suggest several directions for further study First, it would be natural to analyze the transcendentality structure of the vacuum energy of the original PCM as a series in \(B\) and \(1/N\), beyond the maximally transcendental DS sector. Second, it would be interesting to extend our  vacuum-energy analysis to the full transseries in \(1/b\), using the finite-interval WH method, and to study the resulting transcendentality structure. This direction is already hinted at by Borel resummation of the available 35 orders, followed by a convergence-accelerating Richardson transform, which indicates an exponential correction \(\sim e^{-2.00004 b}\), remarkably close to the \(\sim e^{-2b}\) scale expected from the pole structure of \(G_+\). This analysis is in progress and will be reported elsewhere. Third, one may ask whether an analogous DS regime exists for the finite-temperature PCM, and what transcendentality properties it would have. Although such a limit is not currently known, maximal transcendentality could be used in reverse, as a constraint guiding its derivation directly at the level of Q-functions~\cite{Leurent:2015wzw,Kazakov:2010kf}. Finally, it would be interesting to analyze transcendentality patterns for general chemical potentials \cite{Kazakov:2023imu}.

More broadly, the DS PCM offers a new testing ground for number-theoretic structures in strongly coupled QFT. A natural next step is to understand the regularities observed in Sec.~\ref{sec:Observations} from a more conceptual perspective. In particular, it would be interesting to determine whether the observed positivity admits a combinatorial or volume interpretation, and whether the near-closure under odd-zeta derivations points to an underlying motivic organization.

It would also be interesting to search for other maximally or uniformly transcendental models, using the present mechanism as a guide. Our analysis suggests that promising candidates should have WH factorizations with weight-graded small-momentum expansions compatible with the WH recursion. This criterion could in principle be incorporated into recently proposed AI-powered frameworks for discovering new \(R\)/\(S\)-matrices~\cite{Lal:2025nmf,Lal:2023dkj}.

\section*{Materials}

Supplemental Material: Mathematica notebook dspcm.nb, containing the first 22 orders in \(1/b\), calculated directly from \(R_+\), and the first 35 terms of the expansion in \(1/\bar{b}\), calculated purely from \(V\).

\section*{Acknowledgments}
The author thanks Vladimir Kazakov and Konstantin Zarembo for valuable discussions, a careful reading of the manuscript, and related joint work.

\bibliography{MTDSPCM}

\appendix
\section{Small \(b\) expansion}\label{AppendixSmallb}
The kernel \(\mathcal{K}(t)\) has the following small-\(t\) expansion
\begin{gather}
\mathcal{K}(t)=\sum\limits_{m=1}^\infty \frac{2(-1)^m \zeta(2m+1)}{\pi^{2m+2}}t^{2m} \label{Ksmallb}
\end{gather}
The constant \(\delta(b)\) is fixed by the boundary condition \(\epsilon(\pm b)=0\), which gives
\begin{gather}
\epsilon(t)+\int\limits_{-b}^b \left(\mathcal{K}(t-t')-\mathcal{K}(b-t')\right)\epsilon(t')\,dt'=\frac{1}{2}(b^2-t^2)\notag
\end{gather}
Rescaling \(t=bx\) and \(\epsilon(t)=b^2\phi(x)\) and using \eqref{Ksmallb} we rewrite the original equation as
\begin{gather}
\phi(x)=\frac{1-x^2}{2}+\sum\limits_{m=1}^\infty\frac{2(-1)^m \zeta(2m+1)}{\pi^{2m+2}} b^{2m+1}I_m \left[\phi\right],\notag\\
I_m\left[\phi\right]=\int\limits_{-1}^1 \left((1-y)^{2m}-(x-y)^{2m}\right)\phi(y) dy\notag
\end{gather}
This form makes the iterative structure of the equation manifest. Writing 
\begin{gather*}
\phi(x)=\sum\limits_{n=0}^\infty \left(\frac{b}{\pi}\right)^n\phi_n(x),\\
\phi_0(x)=\frac{1-x^2}{2}
\end{gather*}
one gets
\begin{gather}
\phi_n(x)=\sum_{m=1}^{\left[\frac{n-1}{2}\right]}\frac{2(-1)^m \zeta(2m+1)}{\pi}I_m\left[\phi_{n-2m-1}\right]
\end{gather}
This recursion is analogous to Eq.~\eqref{WH_iteration}, but is much simpler. Unlike the latter, however, it does not preserve transcendental weight: each iteration introduces the factor \(\zeta(2m+1)/\pi\).

The first terms in the resulting vacuum energy are
\begin{gather*}
E\underset{b\to 0}{=}\frac{1}{6}\frac{b^{3}}{\pi^{3}}
-\frac{4\zeta(3)}{9\pi}\frac{b^{6}}{\pi^{6}}
+\frac{16\zeta(5)}{15\pi}\frac{b^{8}}{\pi^{8}}
+\frac{32\zeta(3)^2}{27\pi^2}\frac{b^{9}}{\pi^{9}}\\
-\frac{96\zeta(7)}{35\pi}\frac{b^{10}}{\pi^{10}}
-\frac{256\zeta(3)\zeta(5)}{45\pi^2}\frac{b^{11}}{\pi^{11}}
\\
+\left[
\frac{1024\zeta(9)}{135\pi}
-\frac{256\zeta(3)^3}{81\pi^3}
\right]\frac{b^{12}}{\pi^{12}}\\
+\left[
\frac{11264\zeta(5)^2}{1575\pi^2}
+\frac{512\zeta(3)\zeta(7)}{35\pi^2}
\right]\frac{b^{13}}{\pi^{13}}
+\cdots 
\end{gather*}
Finally, we note that an almost identical kernel,
\(K(x)=x[\psi(1+ix)+\psi(1-ix)-2\psi(1)]\), appears in the saddle point integral
equation for the circular \(\mathcal{N}=2\) quiver gauge theory and in its large-\(L\) limit
\cite{Sobko:2025zci}. This leads to a similar weak-coupling expansion for the
Wilson-loop expectation value; see Eq.~(16) of Ref.~\cite{Sobko:2025zci}.

\section{Inverse \(A^{(N)}\)}

\begin{gather}\label{InverseA}
(A^{(N)})^{-1}_{j,s}=
\begin{cases}
0, & s>j, \\[6pt](-1)^{j-N}\,
2^{N+\frac12-j}
, & s=j, \\[14pt]
\dfrac{
2^{N+\frac12-j}
\left(N+\frac12-s\right)
\binom{N-j-\frac12}{j-s-1}
}{
(j-s)\,2^{j-s}
(-1)^{N-j}
},
& s<j .
\end{cases}
\end{gather}

\end{document}